\documentclass[aps, letterpaper, 11pt, onecolumn,
  floatfix, fleqn, notitlepage, endfloats,
  amsmath, amssymb, amsfonts,
  preprintnumbers, showpacs, showkeys]{revtex4-1}\usepackage{graphicx, color}
\makeatletter
\def\maxwidth{ %
  \ifdim\Gin@nat@width>\linewidth
    \linewidth
  \else
    \Gin@nat@width
  \fi
}
\makeatother

\IfFileExists{upquote.sty}{\usepackage{upquote}}{}
\definecolor{fgcolor}{rgb}{0.2, 0.2, 0.2}

\usepackage{framed}
\makeatletter
 {\par\unskip\endMakeFramed%
 \at@end@of@kframe}
\makeatother

\definecolor{shadecolor}{rgb}{.97, .97, .97}
\definecolor{messagecolor}{rgb}{0, 0, 0}
\definecolor{warningcolor}{rgb}{1, 0, 1}
\definecolor{errorcolor}{rgb}{1, 0, 0}
\newenvironment{knitrout}{}{} % an empty environment to be redefined in TeX

\usepackage{alltt}

\usepackage[colorlinks=true, linkcolor=blue, citecolor=blue,
            filecolor=blue, urlcolor=blue]{hyperref}
\usepackage{dcolumn, graphicx, mathptmx, moreverb}
\newcolumntype{.}{D{.}{.}{-1}} % centered on decimal point, including header

\newcommand{\q}[2]{\ensuremath{#1\ \mathrm{#2}}} % physical quantity
\newcommand{\vol}[1]{\textbf{#1}} % volume number

\newcommand{\be}{\begin{equation}}
\newcommand{\ee}{\end{equation}}

\newcommand{\bea}{\begin{eqnarray}}
\newcommand{\eea}{\end{eqnarray}}

\newcommand{\sigmap}{\ensuremath{\sigma_p}} % proton rms size
\newcommand{\sigmaprime}{\ensuremath{\sigma_p'}} % proton rms divergence
\newcommand{\rpipe}{\ensuremath{r_\mathrm{pipe}}} % pipe radius
\newcommand{\kscale}{\ensuremath{k_\mathrm{scale}}} % scale of kicks
\newcommand{\ma}[1]{\ensuremath{\mathbf{#1}}} % matrix

\newcommand{\V}{\ensuremath{\phi}} % electrostatic potential
\newcommand{\IV}{\ensuremath{V}} % integrated potential
\newcommand{\TC}[2]{\ensuremath{T_{#1}\left(\frac{#2}{a}\right)}}
\newcommand{\TP}[2]{\ensuremath{T_{#1}'\left(\frac{#2}{a}\right)}}

\begin{document}

\date{\today}

\title{Calculation of the transverse kicks generated by the bends of a
  hollow electron lens}

\author{Giulio~Stancari}
\email[E-mail: ]{$\langle$stancari@fnal.gov$\rangle$.}

\affiliation{Fermi National Accelerator Laboratory, P.O. Box 500,
  Batavia, Illinois 60510, USA}\thanks{Fermi National Accelerator
  Laboratory (Fermilab) is operated by Fermi Research Alliance, LLC
  under Contract DE-AC02-07\-CH\-11\-359 with the United States
  Department of Energy. This research was supported in part by the US
  DOE LHC Accelerator Research Program (LARP).}

\begin{abstract}
  Electron lenses are pulsed, magnetically confined electron beams
  whose current-density profile is shaped to obtain the desired effect
  on the circulating beam in high-energy accelerators. They were used
  in the Fermilab Tevatron collider for abort-gap clearing, beam-beam
  compensation, and halo scraping. A beam-beam compensation scheme
  based upon electron lenses is currently being implemented in the
  Relativistic Heavy Ion Collider at Brookhaven National
  Laboratory. This work is in support of a conceptual design of hollow
  electron beam scraper for the Large Hadron Collider. It also applies
  to the implementation of nonlinear integrable optics with electron
  lenses in the Integrable Optics Test Accelerator at Fermilab. We
  consider the axial asymmetries of the electron beam caused by the
  bends that are used to inject electrons into the interaction region
  and to extract them. A distribution of electron macroparticles is
  deposited on a discrete grid enclosed in a conducting pipe. The
  electrostatic potential and electric fields are calculated using
  numerical Poisson solvers. The kicks experienced by the circulating
  beam are estimated by integrating the electric fields over straight
  trajectories. These kicks are also provided in the form of
  interpolated analytical symplectic maps for numerical tracking
  simulations, which are needed to estimate the effects of the
  electron lens imperfections on proton lifetimes, emittance growth,
  and dynamic aperture. We outline a general procedure to calculate
  the magnitude of the transverse proton kicks, which can then be
  generalized, if needed, to include further refinements such as the
  space-charge evolution of the electron beam, magnetic fields
  generated by the electron current, and longitudinal proton dynamics.
\end{abstract}

%\pacs{}
%\keywords{}
\preprint{FERMILAB-FN-0972-APC}

\maketitle
\clearpage
\tableofcontents
\clearpage

%%%%%%%%%%%%%%%%%%%%%%%%%%%%%%%%%%%%%%%%%%%%%%%%%%%%%%%%%%%%%%%%%%%%%%%%
\section{Introduction}
%%%%%%%%%%%%%%%%%%%%%%%%%%%%%%%%%%%%%%%%%%%%%%%%%%%%%%%%%%%%%%%%%%%%%%%%

Electron lenses are pulsed, magnetically confined electron beams whose
current-density profile is shaped to obtain the desired effect on the
circulating beam in high-energy accelerators. Electron lenses were
used in the Fermilab Tevatron collider for bunch-by-bunch compensation
of long-range beam-beam tune shifts, for removal of uncaptured
particles in the abort gap, for preliminary experiments on head-on
beam-beam compensation, and for the demonstration of halo scraping
with hollow electron beams. Electron lenses for beam-beam compensation
were built for the Relativistic Heavy Ion Collider at Brookhaven
National Laboratory and are currently being commissioned. Electron
lenses represent one of the most promising options to cope with the
challenges of beam halo scraping and control in the Large Hadron
Collider~\cite{Stancari:PAC:2013, Stancari:TM:2014}. They are also one
of the candidate lattice elements to achieve nonlinear integrable
optics in the Integrable Optics Test Accelerator at
Fermilab~\cite{Nagaitsev:IPAC:2012, Valishev:IPAC:2012}.

\begin{figure}
\includegraphics[width=0.9\textwidth]{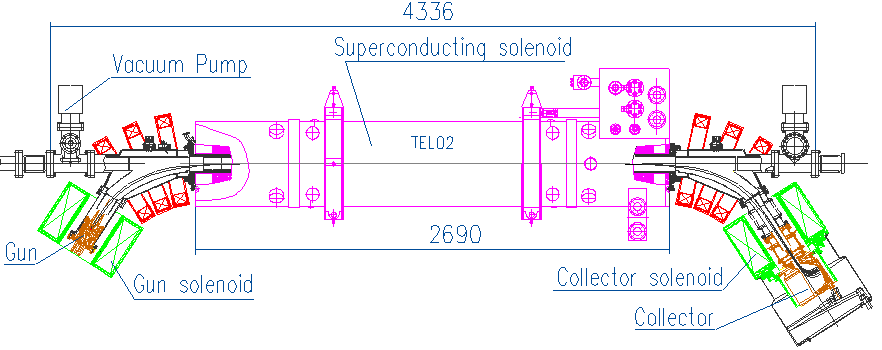}
\caption{Layout of a Tevatron electron lens. The beam is generated in
  the electron gun inside a conventional solenoid and guided by strong
  axial magnetic fields. Inside the superconducting main solenoid, the
  circulating beam interacts with the electromagnetic fields generated
  by the electrons' charge distribution. The electron beam is then
  extracted and deposited in the collector. Dimensions are in
  millimeters.}
\label{fig:TEL-layout}
\end{figure}

Applications of electron lenses often rely on the axial symmetry of
the electrons' current distribution. For instance, in a hollow
electron beam scraper, the halo of the circulating beam is affected by
the electromagnetic fields generated by the
electrons~\cite{Stancari:PRL:2011}. The beam core is unaffected only
if the distribution of the electron charge is axially symmetric. One
cause of asymmetry is the space-charge evolution of the electron beam
as it propagates through the electron lens. In this note, we consider
another effect: the bends that are usually used to inject and extract
the electron beam from the interaction region
(Figure~\ref{fig:TEL-layout}). Although small, these asymmetries may
have detectable effects on core lifetimes and emittances because of
their nonlinear nature, especially when the current of the electron
pulse is changed turn by turn to enhance the halo removal effect by
resonant excitation of selected particles.

%%%%%%%%%%%%%%%%%%%%%%%%%%%%%%%%%%%%%%%%%%%%%%%%%%%%%%%%%%%%%%%%%%%%%%%%
\section{Generation of the electron macroparticles}
%%%%%%%%%%%%%%%%%%%%%%%%%%%%%%%%%%%%%%%%%%%%%%%%%%%%%%%%%%%%%%%%%%%%%%%%

In this note, we focus on the configuration of hollow electron lenses
for the LHC. The parameters of the proton beam are used to calculate
beam sizes in the interaction region. Typical values are reported in
Table~\ref{tab:prot-par}. We assume a round proton beam with an rms
beam size $\sigmap = \q{0.317}{mm}$.

% latex table generated in R 2.15.3 by xtable 1.7-1 package
% Thu Feb 27 17:52:37 2014
\begin{table*}[p]
\begin{ruledtabular}
\caption{Typical proton beam parameters in the interaction
                         region.} 
\label{tab:prot-par}
{\normalsize
\begin{tabular}{ddddd}
 \multicolumn{1}{l}{Kinetic energy} & \multicolumn{1}{l}{Emittance (rms norm.)} & \multicolumn{1}{l}{Lattice function amplitude} & \multicolumn{1}{l}{Size (rms)} & \multicolumn{1}{l}{Divergence (rms)} \\ 
  \multicolumn{1}{c}{$T_p$ [TeV]}
                            & \multicolumn{1}{c}{$\epsilon$ [$\mu$m]}
                            & \multicolumn{1}{c}{$\beta$ [m]}
                            & \multicolumn{1}{c}{\sigmap\ [$\mu$m]}
                            & \multicolumn{1}{c}{\sigmaprime\ [$\mu$rad]} \\ \hline
   7 & 3.75 &  200 &  317 & 1.59 \\ 
  \end{tabular}
}
\end{ruledtabular}
\end{table*}

The hollow electron beam is represented by a cylindrical bent pipe
with a curvature radius $R = \q{0.7}{m}$, inner radius $r_i =
\q{1.268}{mm} = 4 \sigmap$, and outer radius $r_o =
\q{2.386}{mm}$. The ratio between the outer and the inner radius
is chosen to reproduce the dimensions of the cathode in the existing
1-inch hollow electron gun that was built as LHC
prototype~\cite{Li:TM:2012, Moens:Thesis:2013}:
$(\q{12.7}{mm}) / (\q{6.75}{mm})$. The bent tube
of electrons spans a bend angle of $\theta =
\q{30}{deg}$. A total of $N_e = 1048576$ electron
macroparticles is used to reproduce the static charge density
distribution of a $1$-A, $5$-keV electron
beam.

\begin{figure}
\includegraphics[width=0.67\textwidth]{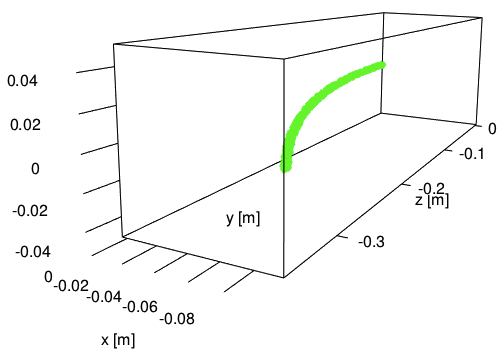}

\caption{The distribution of electron macroparticles mimicks a section
  of the hollow electron beam. (At this magnification, the hole around
  the curved electron beam axis is not visible.)}
\label{fig:electrons}
\end{figure}

A subsample of the distribution of macroparticles is shown in
Figure~\ref{fig:electrons}. The $z$-axis is chosen along the direction
of the circulating proton beam. The $x$-axis points horizontally
outward. The upward direction is represented by the $y$-axis. The
origin of the coordinate system coincides with the point where the
axis of the bent electron beam intersects the axis of the circulating
beam.

%%%%%%%%%%%%%%%%%%%%%%%%%%%%%%%%%%%%%%%%%%%%%%%%%%%%%%%%%%%%%%%%%%%%%%%%
\section{Calculation of electrostatic fields}
%%%%%%%%%%%%%%%%%%%%%%%%%%%%%%%%%%%%%%%%%%%%%%%%%%%%%%%%%%%%%%%%%%%%%%%%

\newcommand{\roix}{\ensuremath{|x| \leq 10\sigmap}}
\newcommand{\roiy}{\ensuremath{|y| \leq 10\sigmap}}

\begin{figure}
\begin{knitrout}\small
\definecolor{shadecolor}{rgb}{0.969, 0.969, 0.969}\color{fgcolor}
\includegraphics[width=\maxwidth]{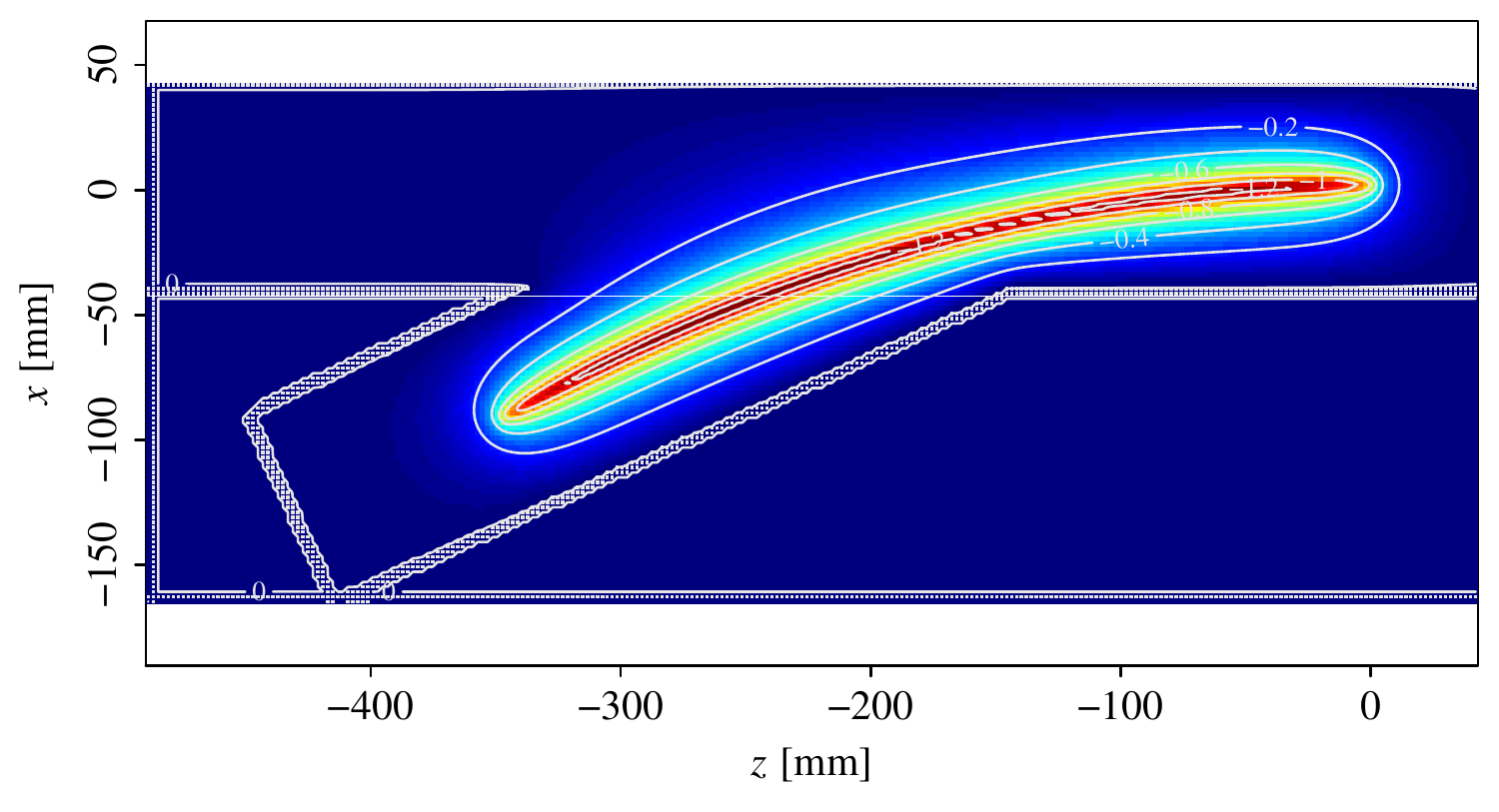} 

\end{knitrout}

\caption{Conductor geometry and calculated electrostatic potential on
  the plane of the bent electron distribution. The contour lines are
  marked with the potential in kilovolts.}
\label{fig:potential-on-plane}
\end{figure}

\begin{figure}
\begin{knitrout}\small
\definecolor{shadecolor}{rgb}{0.969, 0.969, 0.969}\color{fgcolor}
\includegraphics[width=\maxwidth]{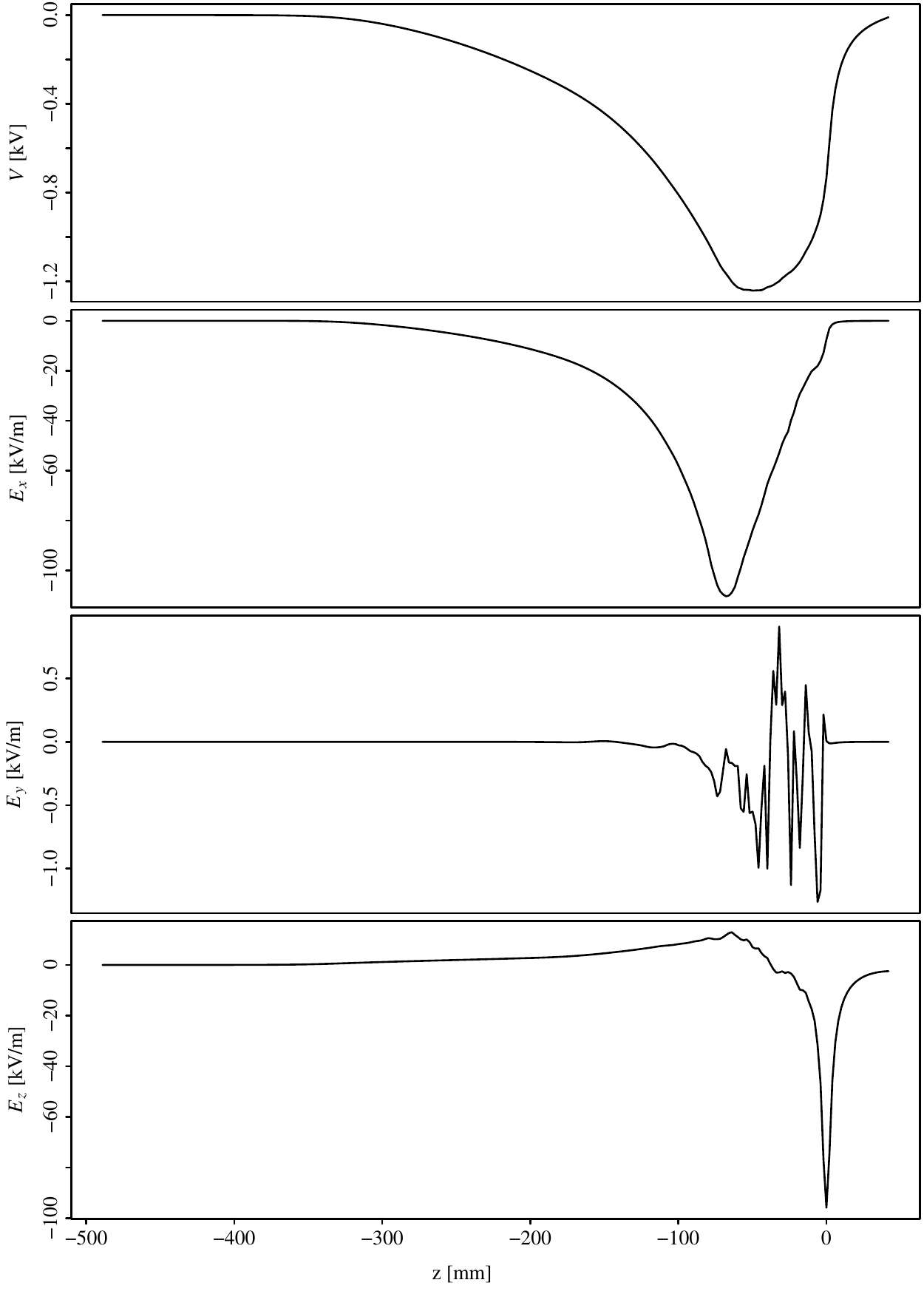} 

\end{knitrout}

\caption{Calculated electrostatic fields on the axis of the proton
  beam ($x=0$, $y=0$).}
\label{fig:fields-on-axis}
\end{figure}

A multigrid Poisson solver within the Warp particle-in-cell
code~\cite{Vay:CSD:2012} is used to calculate the electrostatic
fields. The Dirichlet boundary conditions are defined by a long
cylindrical main beam pipe and by a cylindrical injection beam pipe
stub, joining the main pipe at an angle
(Figure~\ref{fig:potential-on-plane}). This arrangement is a
simplified version of the injection scheme in the Tevatron electron
lenses (Figure~\ref{fig:TEL-layout}). Both pipes have an inner radius
$\rpipe = \q{40}{mm}$. The injection pipe is centered
on the midpoint $(x_C, y_C=0, z_C)$ along the electron beam arc
between the starting edge of the electron beam and the intersection
point between the curved electron beam axis and the main pipe. The
axis of the injection pipe is tangent to the electron beam axis at
$(x_C, y_C=0, z_C)$.

Space is subdivided into 2 discrete grids, a coarse one and a fine
one. The coarse mesh has a spacing of \q{  2}{mm} and covers the whole simulated space, which extends at
least 1~beam pipe radius beyond the charge distribution of the
system. The fine mesh is constructed around the hypothetical central
trajectory of the circulating proton beam. Its spacing is
\q{0.099}{mm}, or a fraction of~\sigmap. It
extends longitudinally for the whole length of the simulated
space. The coverage of the fine mesh is designed to obtain kick maps
for both the core and the halo of the proton beam. We therefore define
a region of interest: \roix\ and \roiy, with $10 \sigmap
= \q{3.17}{mm}$. The fine mesh
extends beyond the region of interest to smoothly connect with the
coarse mesh.

Numerical simulations are run on Linux machines. A full calculation
takes a few minutes on a single processor core. For the coarse mesh,
all the relevant data is written to file: mesh coordinates, charge
density, electrostatic potential, electric field components. For the
fine mesh, only the integrated fields are saved (as described
below). Data analysis, scripting, and report generation are done with
the multi-platform, open-source statistical software~R~\cite{R:2013}.

Figure~\ref{fig:potential-on-plane} shows the electrostatic potential
on the $xz$-plane of the bend. A section of the conductors is marked
by the 0-V equipotential lines and by the grid points in light
gray. The potential near the center of the electron beam is
\q{-1.2}{kV}.

The electrostatic potential and the components of the electric field
on the $z$-axis (axis of the proton beam) of the coarse mesh are shown
in Figure~\ref{fig:fields-on-axis}. As most of the charge lies on the
$x<0$ side of the $zy$-plane, the $x$-component of the electric field
is negative. Because of symmetry, the vertical component $E_y$ should
vanish. Its small fluctuations show the effect of the discrete
distribution of charge. As expected, the $z$-component of the electric
field is positive for large negative~$z$, whereas it becomes negative
after the protons cross the negative charge distribution.

%%%%%%%%%%%%%%%%%%%%%%%%%%%%%%%%%%%%%%%%%%%%%%%%%%%%%%%%%%%%%%%%%%%%%%%%
\section{Calculation of the transverse proton kicks}
%%%%%%%%%%%%%%%%%%%%%%%%%%%%%%%%%%%%%%%%%%%%%%%%%%%%%%%%%%%%%%%%%%%%%%%%

\begin{figure}
\begin{knitrout}\small
\definecolor{shadecolor}{rgb}{0.969, 0.969, 0.969}\color{fgcolor}
\includegraphics[width=\maxwidth]{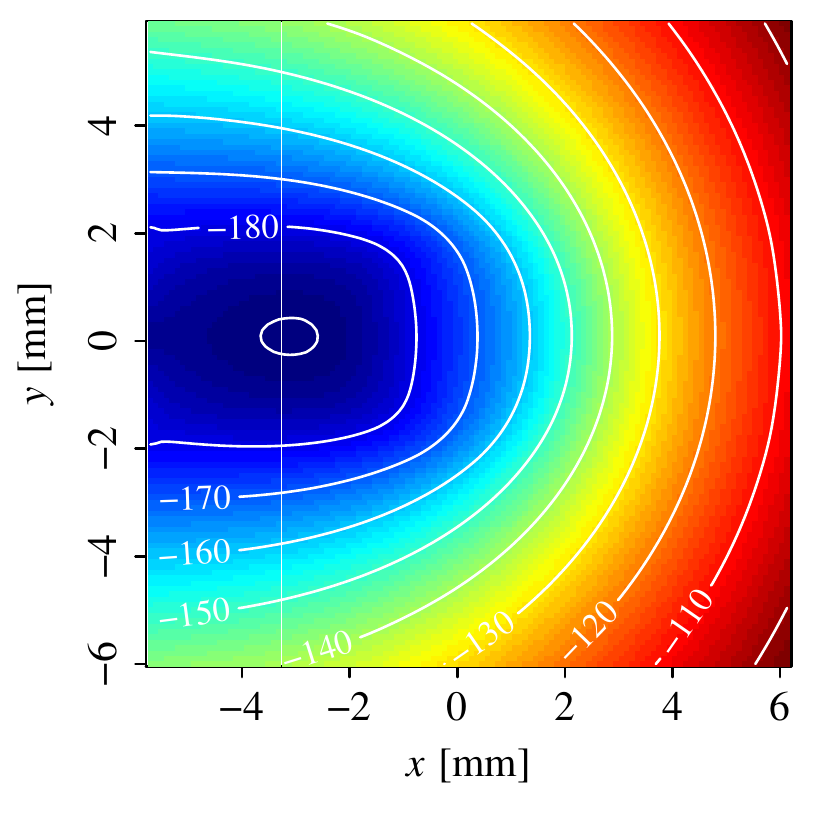} 

\end{knitrout}

\caption{Contour plot of the integrated potential~$\IV(x, y)$. The
  contour lines are labeled in units of $\mathrm{V\cdot m}$.}
\label{fig:integrated-potential-contours}
\end{figure}

\begin{figure}
\begin{knitrout}\small
\definecolor{shadecolor}{rgb}{0.969, 0.969, 0.969}\color{fgcolor}
\includegraphics[width=\maxwidth]{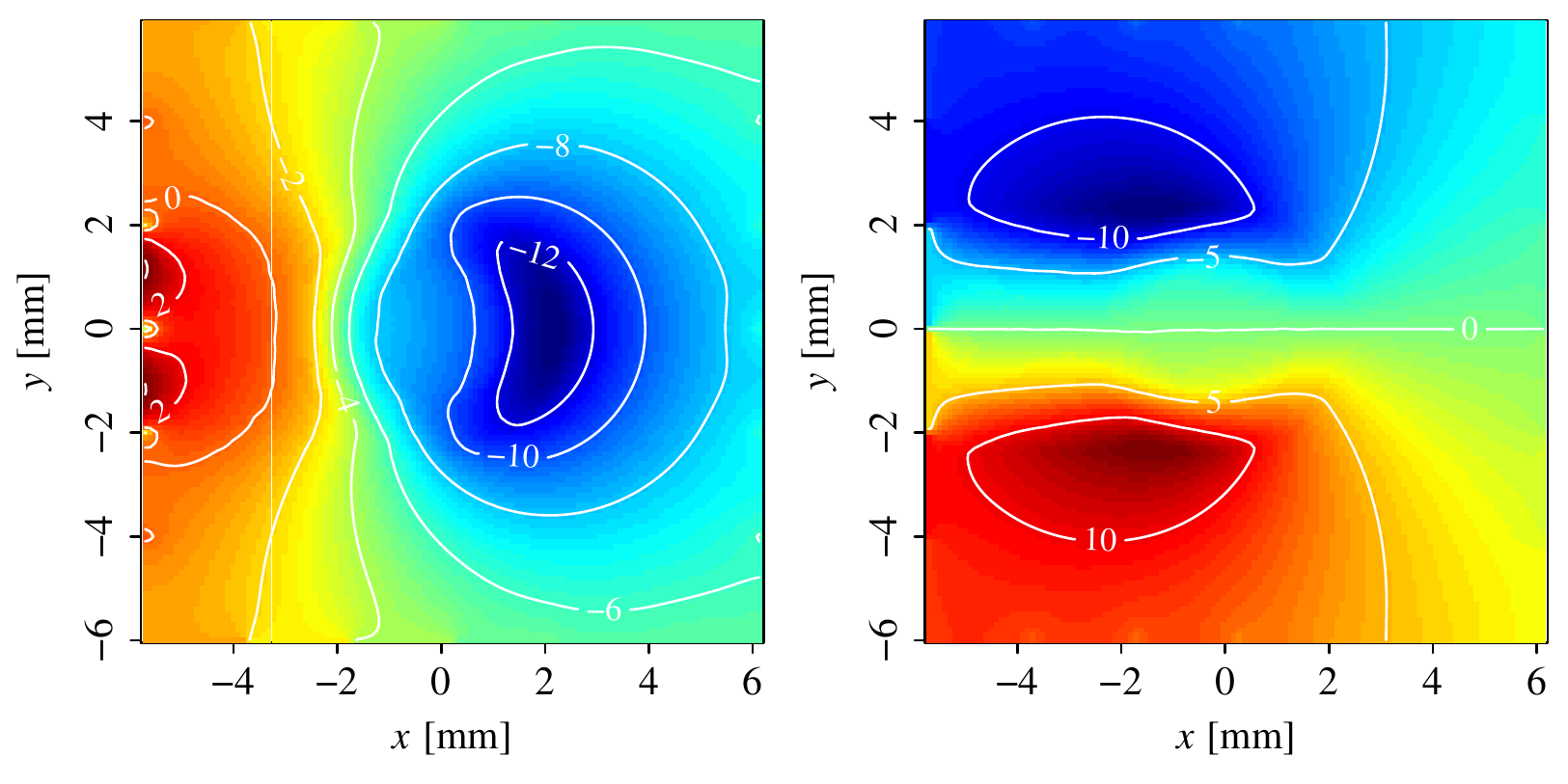} 

\end{knitrout}

\caption{Contour plot of the integrated fields~$k_x(x, y)$ and~$k_y(x,
  y)$. The contour lines are labeled in units of~kV.}
\label{fig:integrated-fields-contours}
\end{figure}

The electrostatic fields are integrated over straight paths to
estimate the kicks on the circulating proton beam. Only transverse
kicks are considered, as the longitudinal kicks of two symmetric bends
cancel out for high-energy protons. The transverse kicks due to two
bends, on the other hand, have the same sign and will add up. The
typical angles of the proton trajectories are of the order of
$\sqrt{\gamma \epsilon} \simeq \sqrt{\epsilon/\beta} =
\q{1.6}{\mu rad}$
(Table~\ref{tab:prot-par}), where $\gamma \equiv (1+\alpha^2)/\beta$
and $\alpha \equiv (-1/2) (d\beta / dz)$ are Courant-Snyder lattice
parameters. The resulting position variations of the protons over the
length of the bend are therefore neglected.

The integrated electrical potential over the proton trajectory,
\be
\IV(x, y) = \int_{z_1}^{z_2} \, \V(x, y, z) \, dz ,
\ee
is shown in Figure~\ref{fig:integrated-potential-contours}.

From Newton's second law, and neglecting magnetic effects, the
transverse kick~$\Delta p$ is related to the impulse of the
electrostatic force~$F$:
\bea
\Delta p_{x, y} & = & \int_{t_1}^{t_2} \, F_{x, y} \, dt =
  \frac{q}{v_z} \int_{z_1}^{z_2} \, E_{x, y} \, dz \label{eq:deltap} \\
\Delta x', \Delta y' & = & \frac{\Delta p_{x, y}}{p_z} =
  \frac{q}{\beta_p^2 \, \gamma_p \, m_p c^2} \int_{z_1}^{z_2} \, E_{x, y} \, dz =
  \frac{1}{(B\rho)_p \cdot v_z} \int_{z_1}^{z_2} \, E_{x, y} \, dz ,
\eea
where $q$, $v$, $\beta_p$, $\gamma_p$, $m_p c^2$, $(B\rho)_p$ are the
charge, velocity, relativistic factors, rest energy, and magnetic
rigidity of the circulating beam, and~$t$ is time. The integrated
electric fields (loosely referred to as `kicks') are defined as
follows:
\bea
k_x(x, y) & \equiv & \int_{z_1}^{z_2} \, E_x(x, y, z) \, dz , \\
k_y(x, y) & \equiv & \int_{z_1}^{z_2} \, E_y(x, y, z) \, dz ,
\eea
and can be applied to particles of different magnetic rigidities.  The
results of the calculation are shown in
Figure~\ref{fig:integrated-fields-contours}. Protons on axis
experience a negative horizontal integrated field of
\q{-9.1}{kV}, whereas the vertical
kick vanishes. For 7-TeV protons, for instance, an
integrated field of 10~kV generates an angular
deviation of 1.4~nrad.

For tracking simulations, one could interpolate the kick maps of
Figure~\ref{fig:integrated-fields-contours} in 2~dimensions for each
particle. Alternatively, for better accuracy and to speed up
calculations, we look for an analytical symplectic parameterization of
the functions $k_x(x, y)$ and~$k_y(x, y)$.

%%%%%%%%%%%%%%%%%%%%%%%%%%%%%%%%%%%%%%%%%%%%%%%%%%%%%%%%%%%%%%%%%%%%%%%%
\section{Parameterization of the transverse kicks}
%%%%%%%%%%%%%%%%%%%%%%%%%%%%%%%%%%%%%%%%%%%%%%%%%%%%%%%%%%%%%%%%%%%%%%%%

The integrated potential is parameterized in terms of the tensor product of
Chebyshev polynomials of the first kind up to
order~$N$~\cite{Press:NR:2007}:
\bea
\IV(x, y) & = & C_{0 0} + C_{1 0} \cdot \TC{1}{x} + C_{0 1}\cdot
\TC{1}{y} + \nonumber \\
 & &   C_{2 0} \cdot \TC{2}{x} + C_{1 1}\cdot \TC{1}{x} \cdot \TC{1}{y} + C_{0
   2} \cdot \TC{2}{y} + \ldots \nonumber \\
 & = & \sum_{n=0}^{N} \sum_{j=0}^{n}
       C_{j,\, (n-j)}\cdot \TC{j}{x} \cdot \TC{n-j}{y}  \label{eq:poly-model}
\eea
where $C_{j l}$ are $(N+1) (N+2) / 2$ dimensional coefficients (in units of
$\mathrm{V\cdot m}$, for instance) and $T_n(u)$ are
Chebyshev polynomials of order~$n$ in the variable~$u$:
\bea
T_0(u) & = & 1 \\
T_1(u) & = & u \nonumber \\
T_n(u) & = & 2 u \cdot T_{n-1}(u) - T_{n-2}(u) . \nonumber
\eea
This parameterization is chosen because, in the interval $u \in [-1,
1]$, Chebyshev polynomials are orthogonal, making the coefficients
independent of each other, and bounded, $T_n(u) \in [-1, 1]$, making
the coefficients of the same order of magnitude. (Because of the
nonzero electron charge density on the proton trajectory, the electric
potential does not satisfy Laplace's equation and it cannot be derived
from a complex harmonic potential decomposable as a superposition of
multipoles.)  The coordinates~$x$ and~$y$ are expressed in units of an
arbitrary length scale ($a = \q{   3.2}{mm}$
for our region of interest), to ensure that the argument of the
polynomials is contained within the unit interval.  Because of the
symmetry of the charge distribution, the potential should be an even
function of~$y$ and only contain even powers of the vertical
coordinate. We therefore require
\be
C_{j 1} = C_{j 3} = C_{j 5} = \ldots = 0
\label{eq:symmetry}
\ee
for any order~$j$ of the polynomials in~$x$.

The kicks are obtained by derivation of the integrated potential:
\bea
k_x(x, y) & = & -\frac{\partial\IV}{\partial x} =
  -\frac{1}{a} \sum_{n=0}^{N} \sum_{j=0}^{n}
       C_{j,\, (n-j)}\cdot \TP{j}{x} \cdot
       \TC{n-j}{y} \label{eq:poly-kicks} \\
k_y(x, y) & = & -\frac{\partial\IV}{\partial y} =
  -\frac{1}{a} \sum_{n=0}^{N} \sum_{j=0}^{n}
       C_{j,\, (n-j)}\cdot \TC{j}{x} \cdot \TP{n-j}{y} \nonumber .
\eea
The derivatives of the Chebyshev polynomials~$T'_n(u)$ can be
calculated recursively as follows:
\be
(1-u^2) \cdot T'_n(u) = n \cdot \left[ T_{n-1}(u) - u \cdot T_n(u) \right],
\ee
or from their trigonometric representation:
\be
T'_n(u) = \frac{n \cdot \sin\left[n \cdot \arccos(u)\right]}{\sqrt{1-u^2}} .
\ee
The cross derivatives of the kicks are
\be
\frac{\partial k_x}{\partial y} = \frac{\partial k_y}{\partial x} =
-\frac{1}{a^2} \sum_{n=0}^{N} \sum_{j=0}^{n}
       C_{j,\, (n-j)}\cdot \TP{j}{x} \cdot \TP{n-j}{y}.
\label{eq:cross-deriv}
\ee
This ensures that the~$z$ component of the curl of the integrated electric field
is zero. This is a necessary and sufficient condition for the $4\times
4$ Jacobian matrix~$\ma{J}$ of initial and final transverse
coordinates $(x^i, y^i; x^f, y^f)$ and momenta $(p_x^i, p_y^i; p_x^f,
p_y^f)$
\be
\ma{J} \equiv
\begin{pmatrix}
  \partial x^f/\partial x^i &
  \partial x^f/\partial p_x^i &
  \partial x^f/\partial y^i &
  \partial x^f/\partial p_y^i \\
  \partial p_x^f/\partial x^i &
  \partial p_x^f/\partial p_x^i &
  \partial p_x^f/\partial y^i &
  \partial p_x^f/\partial p_y^i \\
  \partial y^f/\partial x^i &
  \partial y^f/\partial p_x^i &
  \partial y^f/\partial y^i &
  \partial y^f/\partial p_y^i \\
  \partial p_y^f/\partial x^i &
  \partial p_y^f/\partial p_x^i &
  \partial p_y^f/\partial y^i &
  \partial p_y^f/\partial p_y^i \\
\end{pmatrix}
\equiv
\begin{pmatrix}
1 & 0 & 0 & 0 \\
q & 1 & r & 0 \\
0 & 0 & 1 & 0 \\
s & 0 & t & 1
\end{pmatrix}
\ee
to be symplectic:
\be
\ma{J}^T \, \ma{S} \, \ma{J} - \ma{S}  =
\begin{pmatrix}
0     &  0 & (s-r) & 0 \\
0     &  0 & 0     & 0 \\
(r-s) &  0 & 0     & 0 \\
0     &  0 & 0     & 0
\end{pmatrix} , \mathrm{\ with\ }
\ma{S} \equiv
\begin{pmatrix}
0 & -1 & 0 & 0 \\
1 &  0 & 0 & 0 \\
0 &  0 & 0 & -1 \\
0 &  0 & 1 & 0
\end{pmatrix} ,
\ee
because $s = r$ at any order of approximation.

In the region of interest, we have $M =
4032$ calculated points, each
represented by the coordinates $\left[x_i, y_i\right]$ and fields
$\left[\IV_i, (k_x)_i, (k_y)_i\right]$, with $i = 1, \ldots, M$. The
polynomial model is applied simultaneously (i.e., with the same
coefficients) to the integrated potential and to the kicks. Fitting only
the integrated potential would introduce spurious effects in its
derivatives as the order increases, whereas a fit to only the kicks
cannot determine~$C_{0 0}$. The relative weights of $V$ and $k_{x,y}$
are set according to their ranges in the region of interest:
\bea
V_w & \equiv & \q{62.8}{V\cdot
  m} \label{eq:Vkscale} \\
k_w & \equiv & \q{26.9}{kV} \nonumber .
\eea
The model is linear in the
parameters, and the least-squares coefficients~\ma{C} of this
overdetermined system can be calculated by matrix inversion using
singular-value decomposition (SVD) of the Vandermonde matrices of
powers~\ma{U}~\cite{Press:NR:2007}:
\be
\ma{U} \cdot \ma{C} = \ma{\IV} .
\ee
Explicitly, we have
\newcommand{\fV}{\ensuremath{\upsilon}}
\newcommand{\fk}{\ensuremath{\kappa}}
\newcommand{\fVdef}{\ensuremath{\frac{1}{V_w}}}
\newcommand{\fkdef}{\ensuremath{-\frac{1}{a\cdot k_w}}}
\be
\begin{pmatrix}
  \fV & \fV\cdot\TC{1}{x_1} & \fV\cdot\TC{2}{x_1} & \fV\cdot\TC{2}{y_1} & \fV\cdot\TC{3}{x_1} & \fV\cdot\TC{1}{x_1}\cdot \TC{2}{y_1} & \ldots \\
  \vdots &     &          &          &          & \vdots   &   \\
  \fV & \fV\cdot\TC{1}{x_M} & \fV\cdot\TC{2}{x_M} &
  \fV\cdot\TC{2}{y_M} & \fV\cdot\TC{3}{x_M} & \fV\cdot\TC{1}{x_M}\cdot
  \TC{2}{y_M} & \ldots \\

  0 & \fk\cdot\TP{1}{x_1} & \fk\cdot\TP{2}{x_1} & \fk\cdot\TC{2}{y_1} & \fk\cdot\TP{3}{x_1} & \fk\cdot\TP{1}{x_1}\cdot \TC{2}{y_1} & \ldots \\
  \vdots &     &          &          &          &  \vdots   &   \\
  0 & \fk\cdot\TP{1}{x_M} & \fk\cdot\TP{2}{x_M} &
  \fk\cdot\TC{2}{y_M} & \fk\cdot\TP{3}{x_M} & \fk\cdot\TP{1}{x_M}\cdot
  \TC{2}{y_M} & \ldots \\

  0 & \fk\cdot\TC{1}{x_1} & \fk\cdot\TC{2}{x_1} & \fk\cdot\TP{2}{y_1} & \fk\cdot\TC{3}{x_1} & \fk\cdot\TC{1}{x_1}\cdot \TP{2}{y_1} & \ldots \\
  \vdots &     &          &          &          &  \vdots   &   \\
  0 & \fk\cdot\TC{1}{x_M} & \fk\cdot\TC{2}{x_M} & \fk\cdot\TP{2}{y_M} & \fk\cdot\TC{3}{x_M} & \fk\cdot\TC{1}{x_M}\cdot \TP{2}{y_M} & \ldots
\end{pmatrix}
\cdot
\begin{pmatrix}
  C_{0 0} \\ C_{1 0} \\ C_{2 0} \\ C_{0 2} \\ C_{3 0} \\ C_{1 2} \\ \vdots
\end{pmatrix}
  =
\begin{pmatrix}
    \IV_1/V_w \\  \vdots \\ \IV_M/V_w \\
    (k_x)_1/k_w \\ \vdots \\ (k_x)_M/k_w \\
    (k_y)_1/k_w \\ \vdots \\ (k_y)_M/k_w
\end{pmatrix},
\ee
where the factors \fV\ and \fk\ are defined according to
Eqs.~\ref{eq:poly-model}, \ref{eq:poly-kicks}, and~\ref{eq:Vkscale}:
\bea
\fV & \equiv & \fVdef \\
\fk & \equiv & \fkdef \nonumber .
\eea
Once the coefficients~\ma{C}
are found, the fitted values~\ma{F} and the residuals~\ma{R} are
calculated as follows:
\bea
\ma{F} & = & \ma{U} \cdot \ma{C} \\
\ma{R} & = & \ma{\IV} - \ma{F} .
\eea
An estimate of the random uncertainties on the coefficients is also
provided by this procedure.

\begin{figure}
\begin{knitrout}\small
\definecolor{shadecolor}{rgb}{0.969, 0.969, 0.969}\color{fgcolor}
\includegraphics[width=\maxwidth]{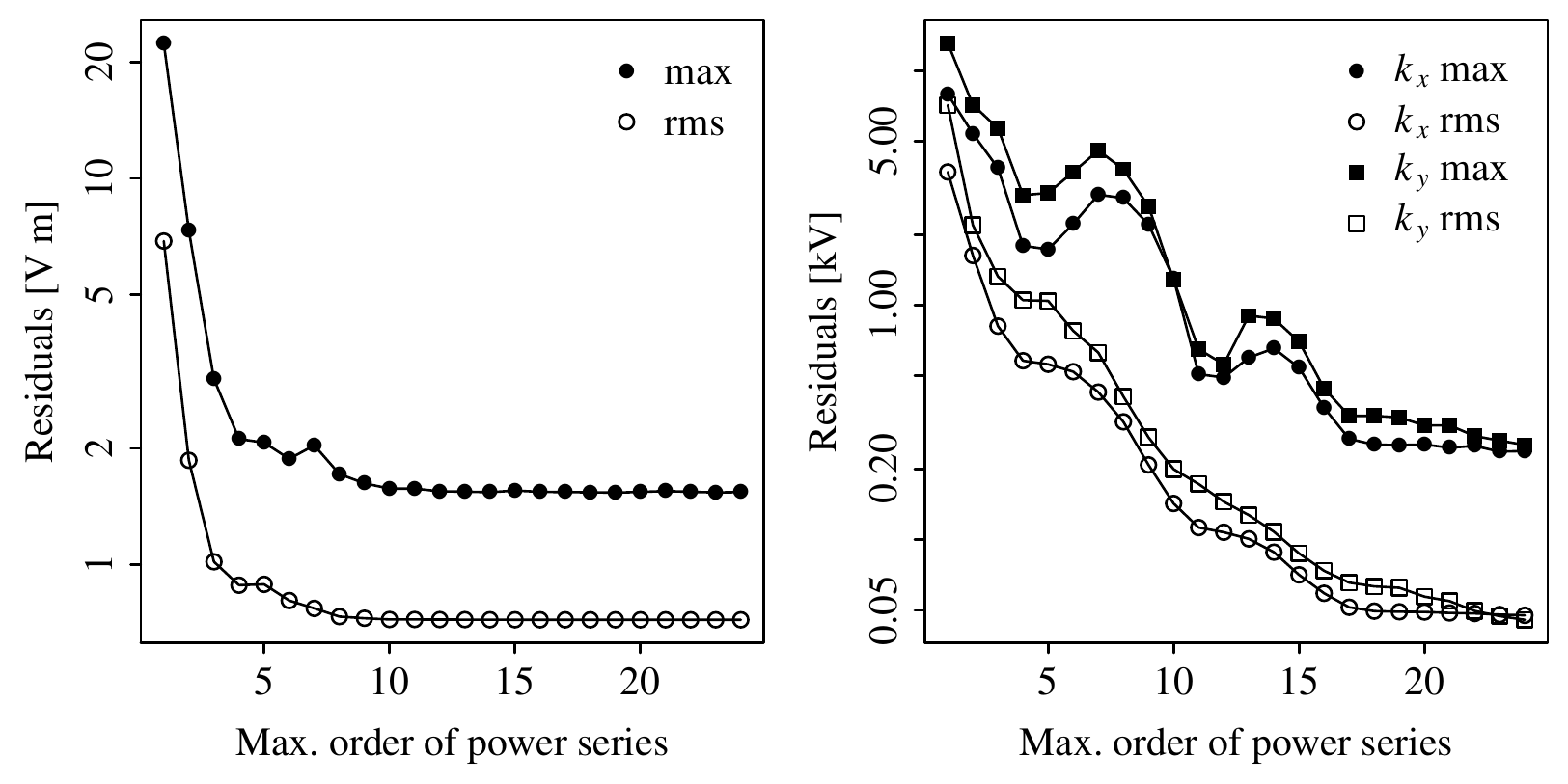} 

\end{knitrout}

\caption{Goodness of fit vs.\ order of the Chebyshev polynomial
  model. The maximum residuals and the standard deviations of the
  residuals are plotted as a function of the maximum order of the
  power series: integrated potential~\IV (left); kicks~$k_x$ and~$k_y$
  (right).}
\label{fig:fit-vs-order}
\end{figure}

\begin{figure}
\begin{knitrout}\small
\definecolor{shadecolor}{rgb}{0.969, 0.969, 0.969}\color{fgcolor}
\includegraphics[width=\maxwidth]{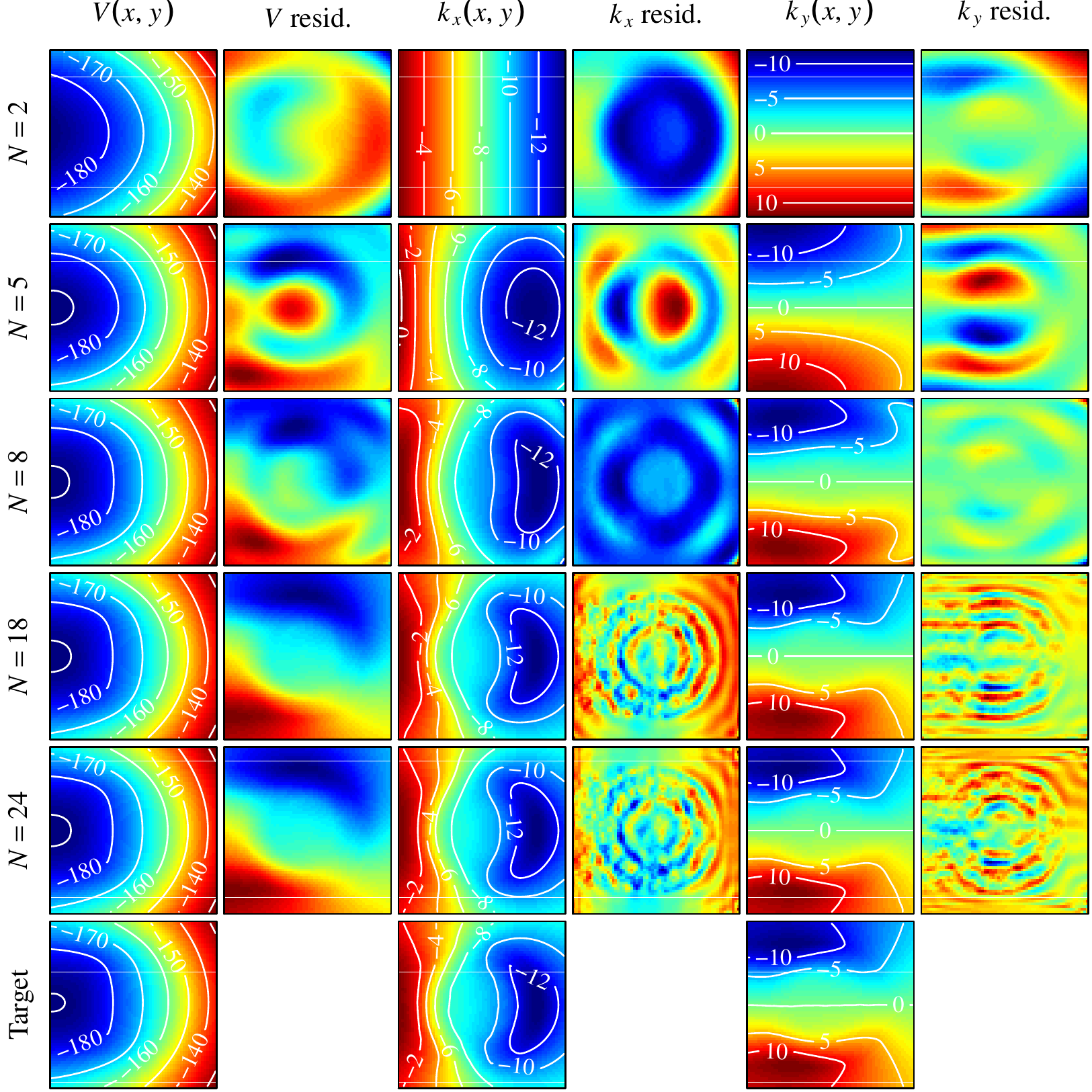} 

\end{knitrout}

\caption{Successive polynomial reconstructions of the integrated
  potential $\IV(x, y)$ (first column) and kicks $k_x(x, y)$, $k_y(x,
  y)$ (third and fifth columns) and the corresponding residuals (even
  columns). Contour line labels are in units of $\mathrm{V\cdot m}$
  for the integrated potential and of~kV for the kicks. For clarity,
  only the pattern of the residuals is shown; their magnitude can be
  inferred from Figure~\ref{fig:fit-vs-order}. The number~$N$ is the
  maximum order of the polynomial. The first 5~rows
  show the polynomial approximations. The bottom row shows the target
  maps (i.e., the Warp calculation).}
\label{fig:convergence-vs-order}
\end{figure}

\begin{figure}
\begin{knitrout}\small
\definecolor{shadecolor}{rgb}{0.969, 0.969, 0.969}\color{fgcolor}
\includegraphics[width=\maxwidth]{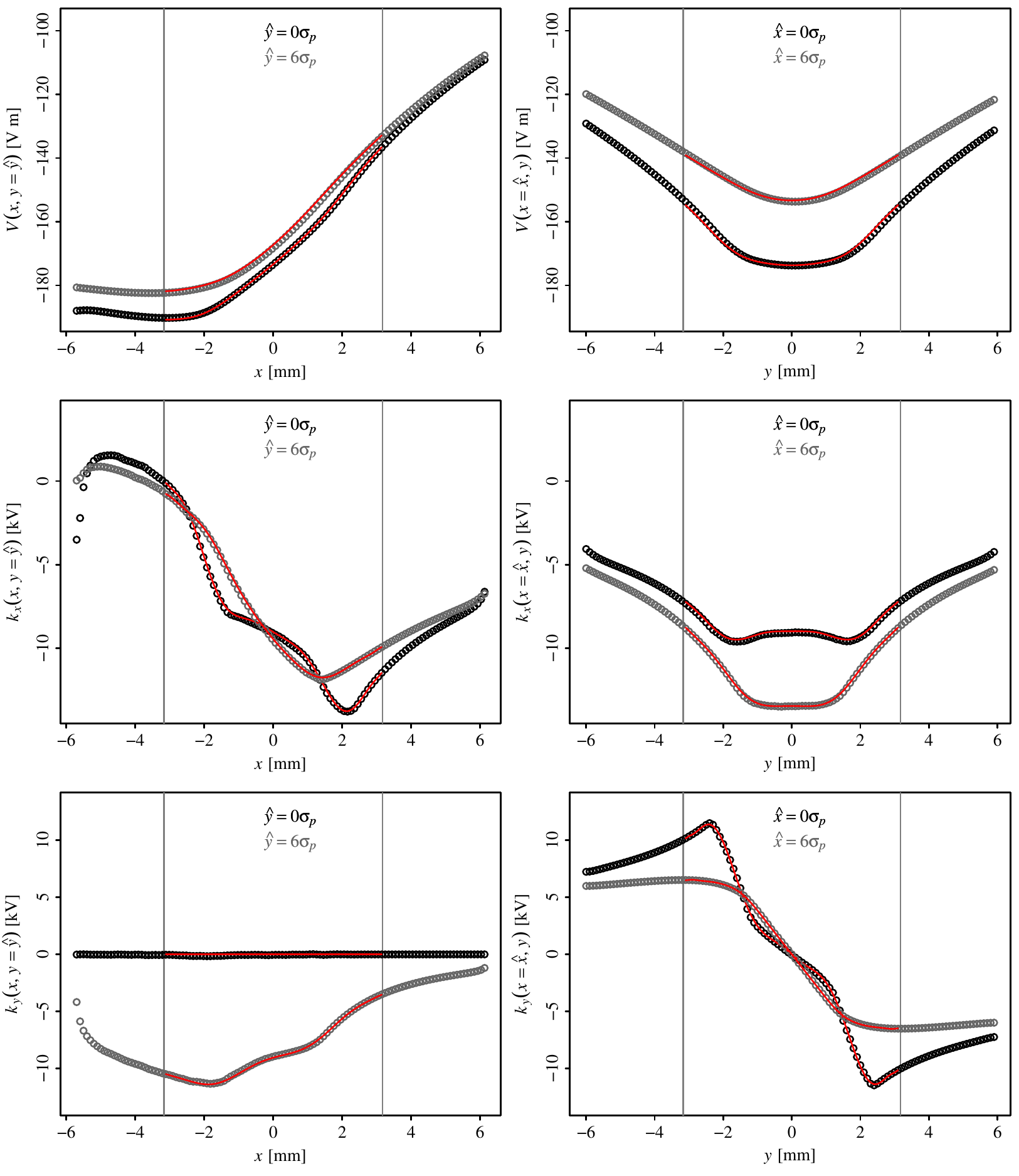} 

\end{knitrout}

\caption{One-dimensional dependence of the integrated potential~$\IV$
  and of the kicks~$k_x$ and~$k_y$ on the transverse coordinates $x$
  and $y$, for $2$~sample values of the other
  coordinate. The empty circles represent the Warp field calculation
  on the fine mesh, whereas the lines represent the polynomial
  interpolation up to order $N = 18$. The vertical gray
  lines represent the region of interest, \roix\ and \roiy.}
\label{fig:fit-sample}
\end{figure}

% latex table generated in R 2.15.3 by xtable 1.7-1 package
% Thu Feb 27 17:52:47 2014
\begin{table*}[p]
\begin{ruledtabular}
\caption{Polynomial coefficients $C_{j l}$
  in $\mathrm{V\cdot m}$, with $a = \q{     3.2}{mm}$ (Eqs.~\ref{eq:poly-model}, \ref{eq:poly-kicks}, and \ref{eq:cross-deriv}).} 
\label{tab:coeff}
{\footnotesize
\begin{tabular}{rrrrrrrrrrr}
  & \ensuremath{T_{0}(y)} & \ensuremath{T_{2}(y)} & \ensuremath{T_{4}(y)} & \ensuremath{T_{6}(y)} & \ensuremath{T_{8}(y)} & \ensuremath{T_{10}(y)} & \ensuremath{T_{12}(y)} & \ensuremath{T_{14}(y)} & \ensuremath{T_{16}(y)} & \ensuremath{T_{18}(y)} \\ 
  \hline
\ensuremath{T_{0}(x)} & \ensuremath{-1.59}E\ensuremath{2} & \ensuremath{9.51} & \ensuremath{-3.44}E\ensuremath{-1} & \ensuremath{-2.99}E\ensuremath{-1} & \ensuremath{9.85}E\ensuremath{-2} & \ensuremath{1.61}E\ensuremath{-2} & \ensuremath{-1.84}E\ensuremath{-2} & \ensuremath{4.38}E\ensuremath{-3} & \ensuremath{-2.47}E\ensuremath{-3} & \ensuremath{1.77}E\ensuremath{-3} \\ 
  \ensuremath{T_{1}(x)} & \ensuremath{2.46}E\ensuremath{1} & \ensuremath{-4.16} & \ensuremath{2.78}E\ensuremath{-2} & \ensuremath{2.92}E\ensuremath{-1} & \ensuremath{-8.09}E\ensuremath{-2} & \ensuremath{-1.36}E\ensuremath{-2} & \ensuremath{2.03}E\ensuremath{-2} & \ensuremath{-2.37}E\ensuremath{-3} & \ensuremath{4.83}E\ensuremath{-3} & 0 \\ 
  \ensuremath{T_{2}(x)} & \ensuremath{4.42} & \ensuremath{-1.2} & \ensuremath{-1.76}E\ensuremath{-1} & \ensuremath{2.12}E\ensuremath{-1} & \ensuremath{-5.01}E\ensuremath{-2} & \ensuremath{-1.3}E\ensuremath{-2} & \ensuremath{1.26}E\ensuremath{-2} & \ensuremath{-1.2}E\ensuremath{-3} & \ensuremath{3.03}E\ensuremath{-3} & 0 \\ 
  \ensuremath{T_{3}(x)} & \ensuremath{-9.56}E\ensuremath{-1} & \ensuremath{2.24}E\ensuremath{-1} & \ensuremath{2.24}E\ensuremath{-1} & \ensuremath{-1.47}E\ensuremath{-1} & \ensuremath{ 2.7}E\ensuremath{-2} & \ensuremath{1.72}E\ensuremath{-2} & \ensuremath{-8.68}E\ensuremath{-3} & \ensuremath{2.51}E\ensuremath{-3} & 0 & 0 \\ 
  \ensuremath{T_{4}(x)} & \ensuremath{-1.91}E\ensuremath{-1} & \ensuremath{-1.09}E\ensuremath{-1} & \ensuremath{2.01}E\ensuremath{-1} & \ensuremath{-9.67}E\ensuremath{-2} & \ensuremath{4.35}E\ensuremath{-3} & \ensuremath{1.69}E\ensuremath{-2} & \ensuremath{-7.47}E\ensuremath{-3} & \ensuremath{ 6.3}E\ensuremath{-4} & 0 & 0 \\ 
  \ensuremath{T_{5}(x)} & \ensuremath{-1.1}E\ensuremath{-2} & \ensuremath{1.56}E\ensuremath{-1} & \ensuremath{-1.39}E\ensuremath{-1} & \ensuremath{5.17}E\ensuremath{-2} & \ensuremath{1.02}E\ensuremath{-2} & \ensuremath{-1.55}E\ensuremath{-2} & \ensuremath{6.43}E\ensuremath{-3} & 0 & 0 & 0 \\ 
  \ensuremath{T_{6}(x)} & \ensuremath{-4.87}E\ensuremath{-2} & \ensuremath{1.24}E\ensuremath{-1} & \ensuremath{-8.35}E\ensuremath{-2} & \ensuremath{1.63}E\ensuremath{-2} & \ensuremath{1.49}E\ensuremath{-2} & \ensuremath{-1.27}E\ensuremath{-2} & \ensuremath{1.73}E\ensuremath{-3} & 0 & 0 & 0 \\ 
  \ensuremath{T_{7}(x)} & \ensuremath{3.74}E\ensuremath{-2} & \ensuremath{-7.08}E\ensuremath{-2} & \ensuremath{3.95}E\ensuremath{-2} & \ensuremath{4.08}E\ensuremath{-3} & \ensuremath{-1.51}E\ensuremath{-2} & \ensuremath{1.01}E\ensuremath{-2} & 0 & 0 & 0 & 0 \\ 
  \ensuremath{T_{8}(x)} & \ensuremath{1.76}E\ensuremath{-2} & \ensuremath{-3.02}E\ensuremath{-2} & \ensuremath{6.32}E\ensuremath{-3} & \ensuremath{1.31}E\ensuremath{-2} & \ensuremath{-1.41}E\ensuremath{-2} & \ensuremath{4.68}E\ensuremath{-3} & 0 & 0 & 0 & 0 \\ 
  \ensuremath{T_{9}(x)} & \ensuremath{-4.81}E\ensuremath{-3} & \ensuremath{8.56}E\ensuremath{-3} & \ensuremath{6.95}E\ensuremath{-3} & \ensuremath{-1.12}E\ensuremath{-2} & \ensuremath{9.66}E\ensuremath{-3} & 0 & 0 & 0 & 0 & 0 \\ 
  \ensuremath{T_{10}(x)} & \ensuremath{2.18}E\ensuremath{-3} & \ensuremath{-5.51}E\ensuremath{-3} & \ensuremath{ 9.8}E\ensuremath{-3} & \ensuremath{-1.02}E\ensuremath{-2} & \ensuremath{5.39}E\ensuremath{-3} & 0 & 0 & 0 & 0 & 0 \\ 
  \ensuremath{T_{11}(x)} & \ensuremath{-2.26}E\ensuremath{-3} & \ensuremath{7.92}E\ensuremath{-3} & \ensuremath{-6.76}E\ensuremath{-3} & \ensuremath{6.75}E\ensuremath{-3} & 0 & 0 & 0 & 0 & 0 & 0 \\ 
  \ensuremath{T_{12}(x)} & \ensuremath{-3.01}E\ensuremath{-3} & \ensuremath{5.02}E\ensuremath{-3} & \ensuremath{-3.26}E\ensuremath{-3} & \ensuremath{2.03}E\ensuremath{-3} & 0 & 0 & 0 & 0 & 0 & 0 \\ 
  \ensuremath{T_{13}(x)} & \ensuremath{2.27}E\ensuremath{-3} & \ensuremath{-2.71}E\ensuremath{-3} & \ensuremath{2.31}E\ensuremath{-3} & 0 & 0 & 0 & 0 & 0 & 0 & 0 \\ 
  \ensuremath{T_{14}(x)} & \ensuremath{3.36}E\ensuremath{-4} & \ensuremath{-6.75}E\ensuremath{-4} & \ensuremath{-3.96}E\ensuremath{-4} & 0 & 0 & 0 & 0 & 0 & 0 & 0 \\ 
  \ensuremath{T_{15}(x)} & \ensuremath{-6.58}E\ensuremath{-4} & \ensuremath{1.04}E\ensuremath{-3} & 0 & 0 & 0 & 0 & 0 & 0 & 0 & 0 \\ 
  \ensuremath{T_{16}(x)} & \ensuremath{5.42}E\ensuremath{-5} & \ensuremath{ 4.7}E\ensuremath{-5} & 0 & 0 & 0 & 0 & 0 & 0 & 0 & 0 \\ 
  \ensuremath{T_{17}(x)} & \ensuremath{ 6.9}E\ensuremath{-4} & 0 & 0 & 0 & 0 & 0 & 0 & 0 & 0 & 0 \\ 
  \ensuremath{T_{18}(x)} & \ensuremath{-1.99}E\ensuremath{-5} & 0 & 0 & 0 & 0 & 0 & 0 & 0 & 0 & 0 \\ 
  \end{tabular}
}
\end{ruledtabular}
\end{table*}

% latex table generated in R 2.15.3 by xtable 1.7-1 package
% Thu Feb 27 17:52:47 2014
\begin{table*}[p]
\begin{ruledtabular}
\caption{Statistical uncertainties on
                 the polynomial coefficients $C_{j l}$.} 
\label{tab:coeff-s}
{\footnotesize
\begin{tabular}{rdddddddddd}
  & \ensuremath{T_{0}(y)} & \ensuremath{T_{2}(y)} & \ensuremath{T_{4}(y)} & \ensuremath{T_{6}(y)} & \ensuremath{T_{8}(y)} & \ensuremath{T_{10}(y)} & \ensuremath{T_{12}(y)} & \ensuremath{T_{14}(y)} & \ensuremath{T_{16}(y)} & \ensuremath{T_{18}(y)} \\ 
  \hline
\ensuremath{T_{0}(x)} & 0.057 & 0.07 & 0.087 & 0.11 & 0.13 & 0.15 & 0.17 & 0.23 & 0.34 & 1.2 \\ 
  \ensuremath{T_{1}(x)} & 0.058 & 0.076 & 0.099 & 0.12 & 0.13 & 0.16 & 0.2 & 0.27 & 0.47 & 0 \\ 
  \ensuremath{T_{2}(x)} & 0.062 & 0.085 & 0.11 & 0.13 & 0.14 & 0.17 & 0.22 & 0.33 & 1.1 & 0 \\ 
  \ensuremath{T_{3}(x)} & 0.074 & 0.097 & 0.12 & 0.13 & 0.15 & 0.19 & 0.27 & 0.46 & 0 & 0 \\ 
  \ensuremath{T_{4}(x)} & 0.083 & 0.11 & 0.12 & 0.14 & 0.17 & 0.22 & 0.32 & 0.93 & 0 & 0 \\ 
  \ensuremath{T_{5}(x)} & 0.089 & 0.12 & 0.13 & 0.15 & 0.19 & 0.27 & 0.44 & 0 & 0 & 0 \\ 
  \ensuremath{T_{6}(x)} & 0.1 & 0.12 & 0.14 & 0.17 & 0.21 & 0.32 & 0.87 & 0 & 0 & 0 \\ 
  \ensuremath{T_{7}(x)} & 0.11 & 0.13 & 0.15 & 0.18 & 0.25 & 0.42 & 0 & 0 & 0 & 0 \\ 
  \ensuremath{T_{8}(x)} & 0.12 & 0.14 & 0.16 & 0.21 & 0.31 & 0.74 & 0 & 0 & 0 & 0 \\ 
  \ensuremath{T_{9}(x)} & 0.13 & 0.15 & 0.18 & 0.25 & 0.4 & 0 & 0 & 0 & 0 & 0 \\ 
  \ensuremath{T_{10}(x)} & 0.14 & 0.16 & 0.21 & 0.29 & 0.66 & 0 & 0 & 0 & 0 & 0 \\ 
  \ensuremath{T_{11}(x)} & 0.15 & 0.18 & 0.24 & 0.38 & 0 & 0 & 0 & 0 & 0 & 0 \\ 
  \ensuremath{T_{12}(x)} & 0.16 & 0.2 & 0.29 & 0.6 & 0 & 0 & 0 & 0 & 0 & 0 \\ 
  \ensuremath{T_{13}(x)} & 0.17 & 0.24 & 0.37 & 0 & 0 & 0 & 0 & 0 & 0 & 0 \\ 
  \ensuremath{T_{14}(x)} & 0.2 & 0.28 & 0.55 & 0 & 0 & 0 & 0 & 0 & 0 & 0 \\ 
  \ensuremath{T_{15}(x)} & 0.24 & 0.36 & 0 & 0 & 0 & 0 & 0 & 0 & 0 & 0 \\ 
  \ensuremath{T_{16}(x)} & 0.28 & 0.52 & 0 & 0 & 0 & 0 & 0 & 0 & 0 & 0 \\ 
  \ensuremath{T_{17}(x)} & 0.35 & 0 & 0 & 0 & 0 & 0 & 0 & 0 & 0 & 0 \\ 
  \ensuremath{T_{18}(x)} & 0.49 & 0 & 0 & 0 & 0 & 0 & 0 & 0 & 0 & 0 \\ 
  \end{tabular}
}
\end{ruledtabular}
\end{table*}

\begin{figure}
\begin{knitrout}\small
\definecolor{shadecolor}{rgb}{0.969, 0.969, 0.969}\color{fgcolor}
\includegraphics[width=\maxwidth]{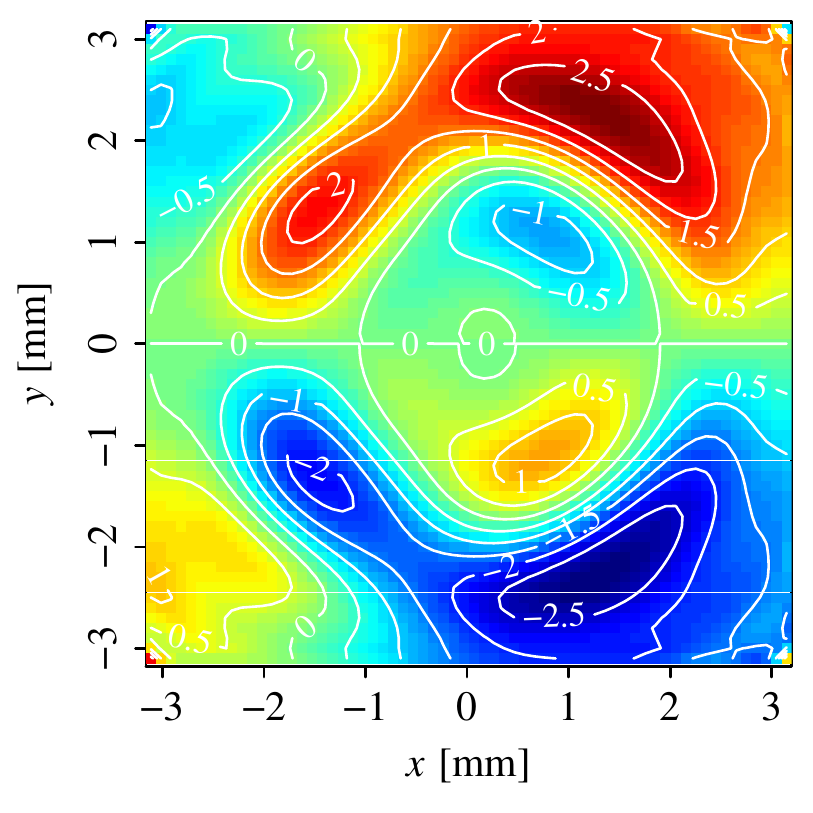} 

\end{knitrout}

\caption{Contour plot of the interpolated cross derivatives $\partial
  k_x/ \partial y =
  \partial k_y/\partial x$ up to order $N = 18$ from
  Eq.~\ref{eq:cross-deriv}. The contour lines are in units of
  $\mathrm{kV/mm}$.}
\label{fig:cross-derivatives}
\end{figure}

Figure~\ref{fig:fit-vs-order} shows how the goodness of fit of the
model (Eqs.~\ref{eq:poly-model}, \ref{eq:symmetry}, and
\ref{eq:poly-kicks}) varies as a function of the maximum order~$N$ of
the power series. The reconstructed integrated potential and the
distribution of residuals as the order of the approximating polynomial
increases are shown in Figure~\ref{fig:convergence-vs-order}. At order
$N = 18$ (our chosen approximation), the standard
deviation of the residuals of the integrated voltage is
\q{0.72}{V\cdot
  m} (compared to a total range of \q{62.8}{V\cdot m}). For the horizontal and vertical kicks, the
standard deviations of residuals are
\q{ 50}{V}
and \q{ 63}{V}, respectively (over a total range of
\q{26.9}{kV}). Inspection of the voltage
residuals (Figure~\ref{fig:convergence-vs-order}), shows a systematic
effect roughly proportional to the vertical derivative of the
integrated potential, suggesting that the potential is not exactly
symmetric with respect to~$y$, but rather shifted by an amount of the
order of the spacing of the fine mesh. This explains why the residuals
do not decrease indefinitely with the order of the polynomial
interpolation (Figure~\ref{fig:fit-vs-order}). This numerical artifact
is negligible for our purposes.

The one-dimensional dependence of the calculated and fitted
functions~$\IV(x, y)$, $k_x(x, y)$, and $k_y(x, y)$ for a few sample
cases is shown in Figure~\ref{fig:fit-sample}. The coefficients of the
order $N = 18$ polynomials and their statistical
uncertainties are reported in Tables~\ref{tab:coeff}
and~\ref{tab:coeff-s}. The interpolated cross derivatives
(Eq.~\ref{eq:cross-deriv}) are shown in
Figure~\ref{fig:cross-derivatives}.

The polynomial expansions of the integrated fields
(Eq.~\ref{eq:poly-kicks}) and the corresponding momentum kicks
(Eq.~\ref{eq:deltap}) are by definition symplectic and can be used in
numerical tracking simulations to estimate their effect on beam
dynamics.

%%%%%%%%%%%%%%%%%%%%%%%%%%%%%%%%%%%%%%%%%%%%%%%%%%%%%%%%%%%%%%%%%%%%%%%%
\section{Dependence of the results on the geometry of the system}
%%%%%%%%%%%%%%%%%%%%%%%%%%%%%%%%%%%%%%%%%%%%%%%%%%%%%%%%%%%%%%%%%%%%%%%%

\begin{figure}
\begin{knitrout}\small
\definecolor{shadecolor}{rgb}{0.969, 0.969, 0.969}\color{fgcolor}
\includegraphics[width=\maxwidth]{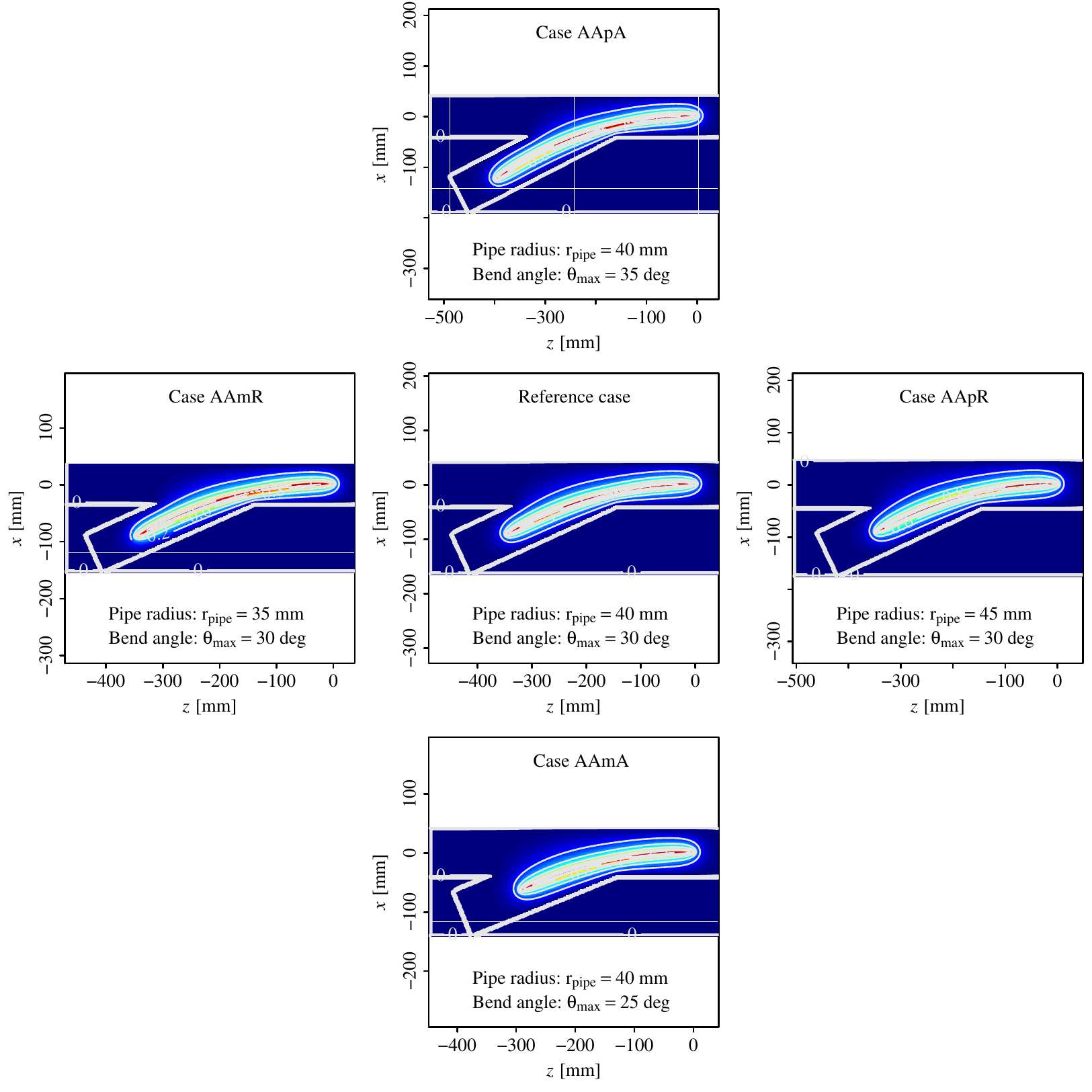} 

\end{knitrout}

\caption{Conductor geometry and potential on the bend plane for each
  simulated case.}
\label{fig:compare-potentials-on-plane}
\end{figure}

To study the sensitivity of the results to the geometry of the system,
cases with different beam pipe radii and injection angles were
simulated. The chosen beam pipe radii were
\q{35}{mm},
\q{40}{mm} (reference case), and
\q{45}{mm}.
These are the inner radii of both the main pipe and the injection pipe.
The chosen bend angles were
\q{25}{deg},
\q{30}{deg} (reference case), and
\q{35}{deg}.
These are the angles spanned by the electron beam.
Figure~\ref{fig:compare-potentials-on-plane} shows the electrostatic
potential on the bend plane ($y=0$) for each of these
5~cases.

\begin{figure}
\begin{knitrout}\small
\definecolor{shadecolor}{rgb}{0.969, 0.969, 0.969}\color{fgcolor}
\includegraphics[width=\maxwidth]{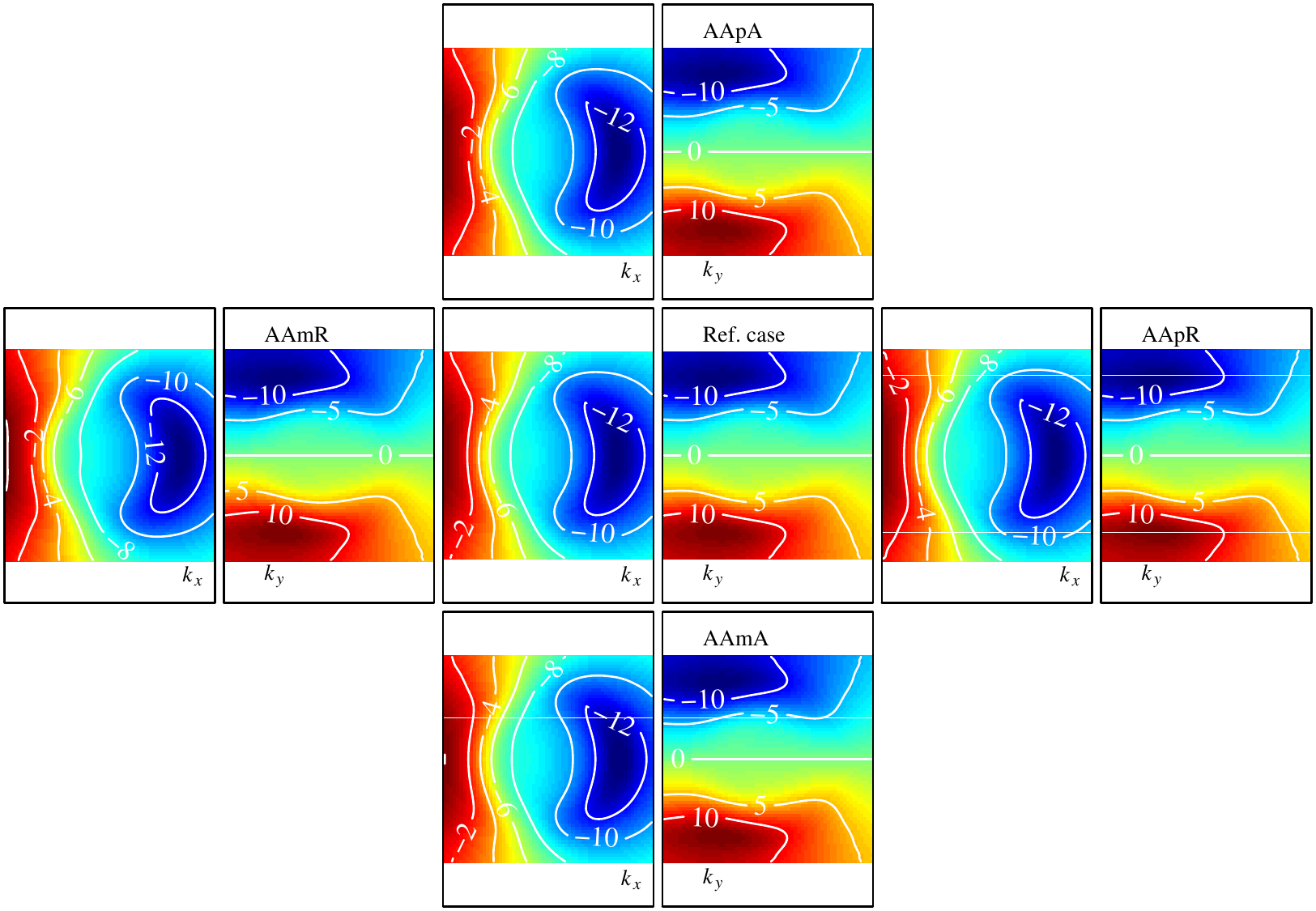} 

\end{knitrout}

\caption{Comparison of the kick maps for the 5 simulated
  cases, in the same order as Figure~\ref{fig:compare-potentials-on-plane}.}
\label{fig:compare-kick-maps}
\end{figure}

\begin{figure}
\begin{knitrout}\small
\definecolor{shadecolor}{rgb}{0.969, 0.969, 0.969}\color{fgcolor}
\includegraphics[width=\maxwidth]{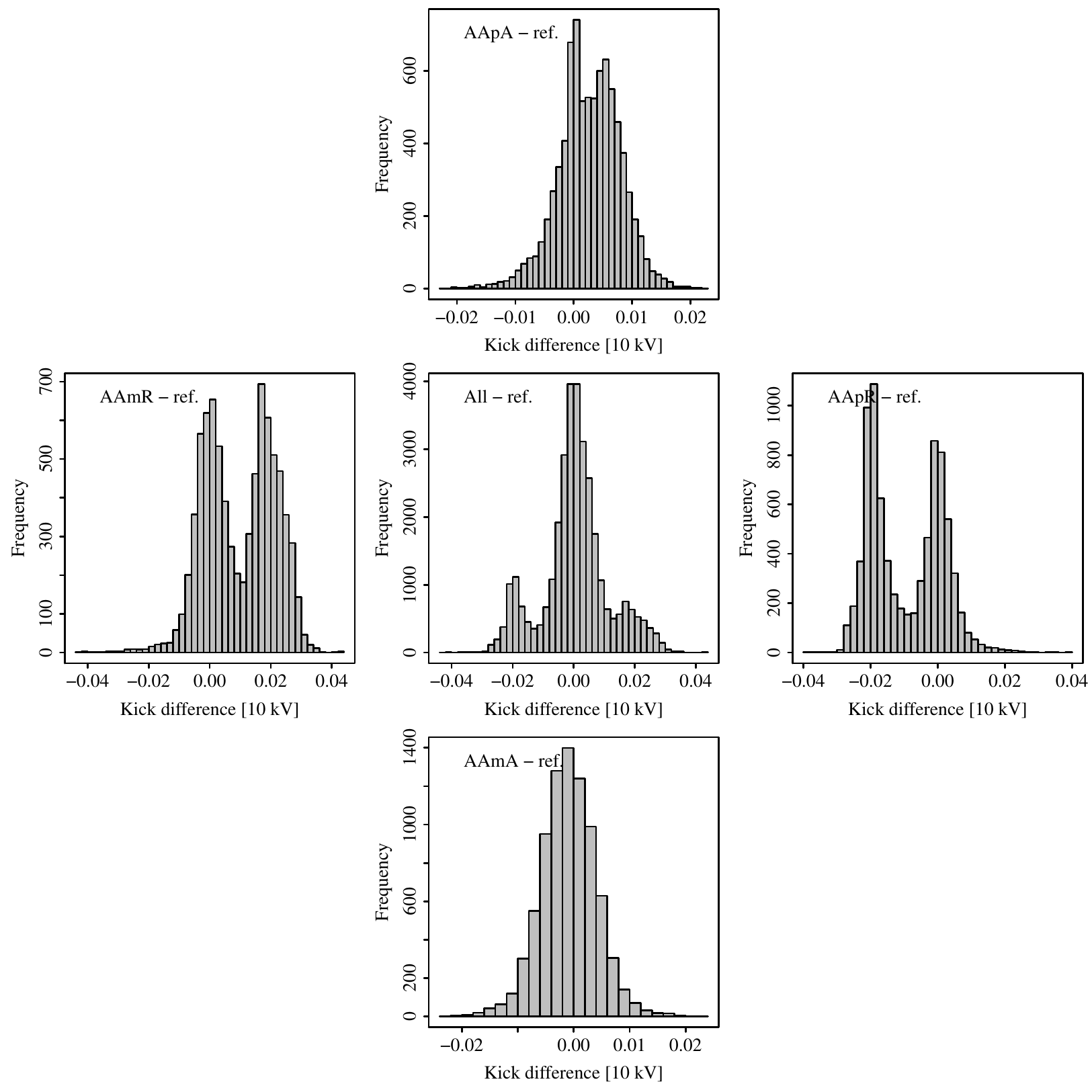} 

\end{knitrout}

\caption{Comparison of the kicks at 4096 random points within
  the region of interest for the 5 simulated cases with
  different geometries. The histograms show the scaled differences
  between the kicks at the corresponding locations. The central
  histogram contains all the differences between the modified cases
  and the reference case.}
\label{fig:case-comparison}
\end{figure}

The calculated kick maps are compared in
Figure~\ref{fig:compare-kick-maps}. They look qualitatively very
similar. A quantitative comparison is presented in
Figure~\ref{fig:case-comparison}. Because the grid points are not
exactly the same in each case, we randomly select 4096 points
in the region of interest (\roix\ and \roiy) and calculate the average
kicks in their neighborhood for each case. The neighborhood is defined
as 2~times the spacing of the fine grid. The histograms of
the differences between the average kicks, divided by the arbitrary
kick scale $\kscale = \q{10}{kV}$, are shown in
Figure~\ref{fig:case-comparison}. The changes in geometry considered
in this study amount to a change in the kicks limited to about
2.3\% (or
0.23~kV).

%%%%%%%%%%%%%%%%%%%%%%%%%%%%%%%%%%%%%%%%%%%%%%%%%%%%%%%%%%%%%%%%%%%%%%%%
\section{Conclusions}
%%%%%%%%%%%%%%%%%%%%%%%%%%%%%%%%%%%%%%%%%%%%%%%%%%%%%%%%%%%%%%%%%%%%%%%%

The electrostatic fields generated by the bends in an electron lens
were calculated. The corresponding symplectic kick maps were provided
as coefficients of truncated power series of orthogonal polynomials
for evaluation with analytical formulas. The goal is to asses some of
the effects of electron-lens asymmetries on the circulating beam in
the case of the proposed proton halo scraper for the LHC. The same
technique can be applied to electron lenses with different
current-density profiles (Gaussian, flat, etc.) to study perturbations
in nonlinear integrable optics for the IOTA ring at Fermilab.

The present work may be further extended by taking into account the
space-charge evolution of the electron beam, the magnetic fields, and
longitudinal proton dynamics.

%%%%%%%%%%%%%%%%%%%%%%%%%%%%%%%%%%%%%%%%%%%%%%%%%%%%%%%%%%%%%%%%%%%%%%%%
% Tables and Figures
%%%%%%%%%%%%%%%%%%%%%%%%%%%%%%%%%%%%%%%%%%%%%%%%%%%%%%%%%%%%%%%%%%%%%%%%

\clearpage
\printtables

\clearpage
\printfigures

\clearpage

%%%%%%%%%%%%%%%%%%%%%%%%%%%%%%%%%%%%%%%%%%%%%%%%%%%%%%%%%%%%%%%%%%%%%%%%

\end{document}